\documentclass[aps,twocolumn,showpacs,amsmath,10pt]{revtex4}

\usepackage{bm}
\usepackage{graphicx}

\newcommand{\ket}[1]{\, | #1 \rangle}

\newcommand{\expv}[1]{\langle #1 \rangle}

\newcommand{\om}{\omega}
\newcommand{\Om}{\Omega}

\newcommand{\de}{\delta}

\newcommand{\hlf}{\mbox{$\frac{1}{2}$}}
\newcommand{\br}{\mathbf{r}}
\newcommand{\ve}{\varepsilon}
\newcommand{\Js}{\mathcal{J}}

\newcommand{\hb}{\hat{b}}
\newcommand{\ha}{\hat{a}}
\newcommand{\hbd}{\hat{b}^{\dagger}}
\newcommand{\had}{\hat{a}^{\dagger}}
\newcommand{\hn}{\hat{n}}

\newcommand{\be}{\begin{equation}}
\newcommand{\ee}{\end{equation}}
\newcommand{\bea}{\begin{eqnarray}}
\newcommand{\eea}{\end{eqnarray}}
\newcommand{\beann}{\begin{eqnarray*}}
\newcommand{\eeann}{\end{eqnarray*}}
\newcommand{\besal}[1]{\begin{subequations}\label{#1}\begin{eqnarray}}
\newcommand{\besa}{\begin{subequations}\begin{eqnarray}}
\newcommand{\eesa}{\end{eqnarray}\end{subequations}}

\begin{document}

\title{Atom number filter in an optical lattice}

\author{Georgios M. Nikolopoulos}
\author{David Petrosyan}
\affiliation{Institute of Electronic Structure and Laser, 
Foundation for Research and Technology -- Hellas, 
71110 Heraklion, Crete, Greece}

\date{\today}

\begin{abstract}
We present an efficient procedure to filter out from an optical lattice,
having inhomogeneous site occupation number, only preselected number 
of bosonic atoms per site and place them into another internal atomic 
state, creating thereby a lattice with desired site occupation number.
\end{abstract}


\maketitle

%\section*{Introduction}

Ultracold atoms in optical lattices \cite{OptLatRev} represent
a remarkably clean and controllable system \cite{optlatMIt,optlatMIe} 
to realize the fundamental Bose-Hubbard model \cite{bosMI}. 
Its two main ingredients are the atom tunneling, or hopping $J$, 
between the neighbouring lattice sites and the on-site atom-atom 
interaction $U$. In a homogeneous lattice, when the kinetic energy 
due to the inter-site hopping dominates, $J \gtrsim U$, the atoms 
are delocalized over the entire lattice yielding a superfluid (SF) phase,
while in the opposite regime of strong on-site interaction, $U \gg J$, 
the hopping is energetically suppressed resulting in a Mott insulator (MI) 
phase with fixed integer number $n$ of localized atoms at each lattice 
site. When a deep optical lattice is superimposed by a shallow confining 
potential, there can be MI phases with occupation numbers of $n =0,1,2,...$
in successive spatial shells \cite{DMLVW,n12345,n12}, separated by 
SF phases with intermediate mean occupation number corresponding to 
delocalized atoms on top of the filled MI shell. 

Experimentally \cite{optlatMIe}, the quantum phase transition between the 
SF and MI phases is implemented by adiabatically increasing the lattice 
depth which results in the reduction of intersite tunneling amplitude 
and simultaneous increase of the on-site interaction \cite{optlatMIt}. 
If, however, the lattice potential is raised quickly, so that the 
tunneling is suddenly switched off, each site occupation ``freezes'' 
to whatever atom-number distribution it corresponded to just before 
the switching off, be it a SF, a MI, or a spatially-dependent 
combination of the two phases.

In this paper, we propose a very efficient method to filter out
from such a frozen ($J =0$) optical lattice only the desired 
number $N$ of atoms per site. This is achieved by using an external 
field which couples the initially populated internal atomic state 
$\ket{a}$ to another internal state $\ket{b}$ trapped by a second 
optical lattice potential. We show that, for strong enough 
state- (or lattice-) dependent on-site interactions, the coupling 
field with properly tuned frequency will selectively transfer 
to the second lattice only the singles ($N=1$), the pairs ($N=2$), 
or the triples ($N=3$) of atoms, via the corresponding $N$-photon 
resonant transition. Hence, after the transfer, the second lattice 
will only have the desired site occupation number $N = 1, 2$, or $3$, 
while the first lattice will contain all the other occupation numbers 
$n \neq N$. 

Before proceeding, we note related, but different, earlier work.
Rabl {\it et al.} \cite{MBHMSR} proposed to reduce the site 
occupation number defects in an optical lattice by adiabatically 
transferring a chosen number of atoms to another internal state.
DeMarco {\it et al.} \cite{DMLVW} studied similar systems 
employing rapid adiabatic transfer of atoms to the second internal 
state, or inducing resonant single-photon Rabi oscillations 
between the atomic states with occupation number--dependent Rabi 
frequencies. Mohring {\it et al.} \cite{RDFCZ} discussed coherent
extraction of atoms from a BEC reservoir into the quantum 
tweezers---tight trap---using adiabatic and resonant transfer 
techniques.

%\section*{The Model}

Considering only two internal atomic states and corresponding optical
lattice potentials, the Hamiltonian of the system takes the form
\bea
H &=& \sum_j \big[ 
(\hbar \om_a + \ve_{a,j}) \hn_{a,j} + \hlf U_{aa} \hn_{a,j} (\hn_{a,j} -1) 
\nonumber \\ & & \quad
+ (\hbar \om_b + \ve_{b,j}) \hn_{b,j} + \hlf U_{bb} \hn_{b,j} (\hn_{b,j} -1)  
\nonumber \\ & & \quad  
+  U_{ab} \hn_{a,j} \hn_{b,j} + \hbar \Om (\hbd_j \ha_j e^{-i \om t} + 
\had_j \hb_j e^{i \om t}) \big] . \quad  \label{Ham} 
\eea
Here $\ha_j$ ($\had_j$) and $\hb_j$ ($\hbd_j$) are the annihilation 
(creation) operators for bosonic atoms in the internal states $\ket{a}$ 
and $\ket{b}$, of energies $\hbar \om_a$ and $\hbar \om_b$, localized 
at lattice site $j$, with single-particle energies $\ve_{a,j}$ and $\ve_{b,j}$, 
and $\hn_{a,j} \equiv \had_j \ha_j$ and $ \hn_{b,j} \equiv \hbd_j \hb_j$ 
are the corresponding number operators. A natural basis for 
Hamiltonian (\ref{Ham}) is that of the eigenstates $\ket{n_{\alpha,j}}$ 
of operators $\hn_{\alpha,j}$ whose eigenvalues $n = 0,1,2,\ldots$ denote 
the number of atoms in the corresponding state $\ket{\alpha}$ ($\alpha = a,b$)
at site $j$. Next, 
$U_{\alpha \alpha} = g_{\alpha \alpha} \int d^3 r |w_{\alpha}(\mathbf{r})|^4$
is the on-site interaction energy for the atoms in state $\ket{\alpha}$, 
and $U_{ab} = g_{ab} \int d^3 r |w_a(\mathbf{r})|^2|w_b(\mathbf{r})|^2$ 
is the interaction between the $\ket{a}$ and $\ket{b}$ atoms, where
$g_{\alpha \alpha'} \equiv 4 \pi a_{\alpha \alpha'} \hbar^2 /M$,  
with $a_{\alpha \alpha'}$ being the corresponding $s$-wave scattering length, 
$M$ the atomic mass and $w_{\alpha}(\mathbf{r})$ the (localized) Wannier 
function of the lowest Bloch band of the corresponding lattice 
potential \cite{OptLatRev,optlatMIt}. 
Finally, $\Om = \Om_{ab} \int d^3 r w_a^*(\mathbf{r}) w_b(\mathbf{r})$
is the coupling amplitude between the localized atoms in states 
$\ket{a}$ and $\ket{b}$, which is induced by an external field 
with the ``bare'' (free--atom) Rabi frequency $\Om_{ab}$. This field can 
be a microwave field of frequency $\om \sim \om_b - \om_a$ coupling the 
atomic hyperfine states $\ket{a}$ and $\ket{b}$ through a magnetic
dipole transition, or an optical bi-chromatic field inducing Raman 
transition $\ket{a} \to \ket{b}$, in which case $\om$ is the frequency 
difference between the two field components (the corresponding 
differential ac Stark shift of $\ket{a}$ and $\ket{b}$ can be 
incorporated in $\om_a$ or $\om_b$). Note that the rotating-wave
approximation, requiring $\Om \ll \om$, is presumed in the last 
term of Eq.~(\ref{Ham}).

In deep optical lattices, the Wannier functions 
$w_{\alpha} (\br - \br_{\alpha,j})$ localized on individual sites $j$ 
can be well approximated \cite{OptLatRev} by the ground-state wavefunction
of a harmonic oscillator centered at $\br_{\alpha,j}$, 
\be
w_{\alpha} (\br - \br_{\alpha,j}) \approx 
\left( \frac{1}{\pi \sigma_{\alpha}^2} \right)^{3/4} 
\exp \left[ - \frac{(\br - \br_{\alpha,j})^2}{2 \sigma_{\alpha}^2} \right] , 
\label{GaussWann}
\ee
where the width $\sigma_{\alpha} = \sqrt{\hbar/M \nu_{\alpha}}$ is expressed 
through the vibrational frequency $\nu_{\alpha} = \sqrt{2 \pi^2 V_{\alpha}/M d^2}$
determined by the lattice potential amplitude $V_{\alpha}$ and period $d$. 
For the interaction parameters of Hamiltonian (\ref{Ham}) we then obtain
\besa
\label{Uaap}
U_{\alpha \alpha} & \simeq & \frac{g_{\alpha \alpha}}
{(2 \pi \sigma_{\alpha}^2 )^{3/2}} \propto a_{\alpha \alpha} V_{\alpha}^{3/4}
\quad (\alpha = a,b) , \\
U_{ab} & \simeq & \frac{g_{ab}} {[\pi (\sigma_{a}^2 + \sigma_{b}^2)]^{3/2}} \, 
\exp \left[-\frac{\de r^2}{\sigma_{a}^2 + \sigma_{b}^2} \right] , \\
\Om & \simeq & \Om_{ab} 
\left( \frac{\sigma_{a} \sigma_{b}}{\sigma_{a}^2 + \sigma_{b}^2} \right)^{3/2} 
\exp \left[-\frac{\de r^2}{2(\sigma_{a}^2 + \sigma_{b}^2)} \right] , 
\eesa
where $\de r \equiv |\br_{a,j} - \br_{b,j}| < d$ is a possible offset 
of the lattice potentials for the atoms in states $\ket{a}$ and $\ket{b}$ 
\cite{FKCSAMetal}. These expressions attest to the controllability of the 
atom-atom interactions $U_{\alpha \alpha'}$ and coupling $\Om$ through the 
interatomic scattering lengths $a_{\alpha \alpha'}$ ($\alpha, \alpha' = a,b$);
the optical lattice parameters, including the lattice modulation depths 
$V_\alpha$, affecting $\sigma_{\alpha}$'s, and the relative offset $\de r$; 
as well as the external coupling field amplitude, affecting $\Om_{ab}$. 
In the experiments, typically 
$U_{\alpha \alpha'}/\hbar \lesssim 2 \pi \times 10\:$kHz 
\cite{OptLatRev,optlatMIe,n12345,n12}. We emphasise that the single-band
approximation inherent in Hamiltonian (\ref{Ham}) requires that the
atom-atom interactions be small compared to the
excited band energies, $U_{\alpha \alpha'} < \hbar \nu_{\alpha}$.

While, in general, $\ve_{a,j}$ and $\ve_{b,j}$ need not be uniform 
throughout the lattice, due to, e.g., shallow external trap, 
we assume that the difference $|\ve_{a,j} - \ve_{b,j}|$ is constant,
and typically small compared to $\om_b - \om_a$, for all $j$. 
Accordingly, we define $\hbar \om_{ba} \equiv \hbar \om_b + \ve_{b,j} 
- (\hbar \om_a + \ve_{a,j}) = \mathrm{const} \; \forall j$ 
and omit the subscript $j$ from now on.

%\section*{Transfer Protocols}

Assume that initially all the atoms are in state $\ket{a}$,
with the sites of the corresponding lattice having arbitrary 
occupations $\ket{n_a}$, $n= 0,1,2,\ldots$, and all the 
sites of the other lattice empty, $\ket{0_b}$. 

Before describing our main idea of atom-number filter, we briefly 
consider a simple but instructive case of uniformly interacting 
$U_{\alpha \alpha'} = U$ ($\alpha, \alpha' = a,b$), or non-interacting 
$U = 0$, atoms subject to a resonant coupling field $\om =\om_{ba}$. 
Within an $N$-atom subspace ($N =n_a + n_b$), the transition matrix 
element of Hamiltonian (\ref{Ham}) between any pair of states of 
the form $\ket{(N-n)_a,n_b}$ and $\ket{(N-n-1)_a,(n+1)_b}$ is given by 
$\Om \sqrt{(N-n)(n+1)}$, as dictated by the bosonic nature of the atoms.
This coupling pattern makes the system formally analogous  to a spin-$\Js$
in a magnetic field. Indeed, as we may recall from the theory of angular 
momentum, with the quantization direction along an axis perpendicular 
to the magnetic field direction, the matrix elements for the transitions 
$\ket{\Js,m} \to \ket{\Js,m+1}$ between the neighboring magnetic  
sub-states ($m = - \Js, \ldots, \Js$) are given by  
$\Om \sqrt{(\Js-m)(\Js+m+1)}$. The spin then exhibits non-dispersive 
precession about the field direction with the Larmor frequency $\Om$. 
Setting formally $\Js = \frac{1}{2} N$ and $m = n - \frac{1}{2}N $ 
leads to the above matrix elements of (\ref{Ham}). A curious consequence 
of this analogy is that, within {\em any} $N$-atom subspace, the resonant 
coupling field will induce oscillations between states $\ket{N_a,0_b}$ 
and $\ket{0_a,N_b}$ with the {\em same} frequency $\Om$. And in particular, 
starting from all the atoms in state $\ket{a}$ and arbitrary site 
occupation numbers of the corresponding lattice, at time $\tau = \pi/2 \Om$, 
all the atoms will simultaneously be transferred to state $\ket{b}$.

\begin{figure*}[t]
\centerline{\includegraphics[width=0.7\textwidth]{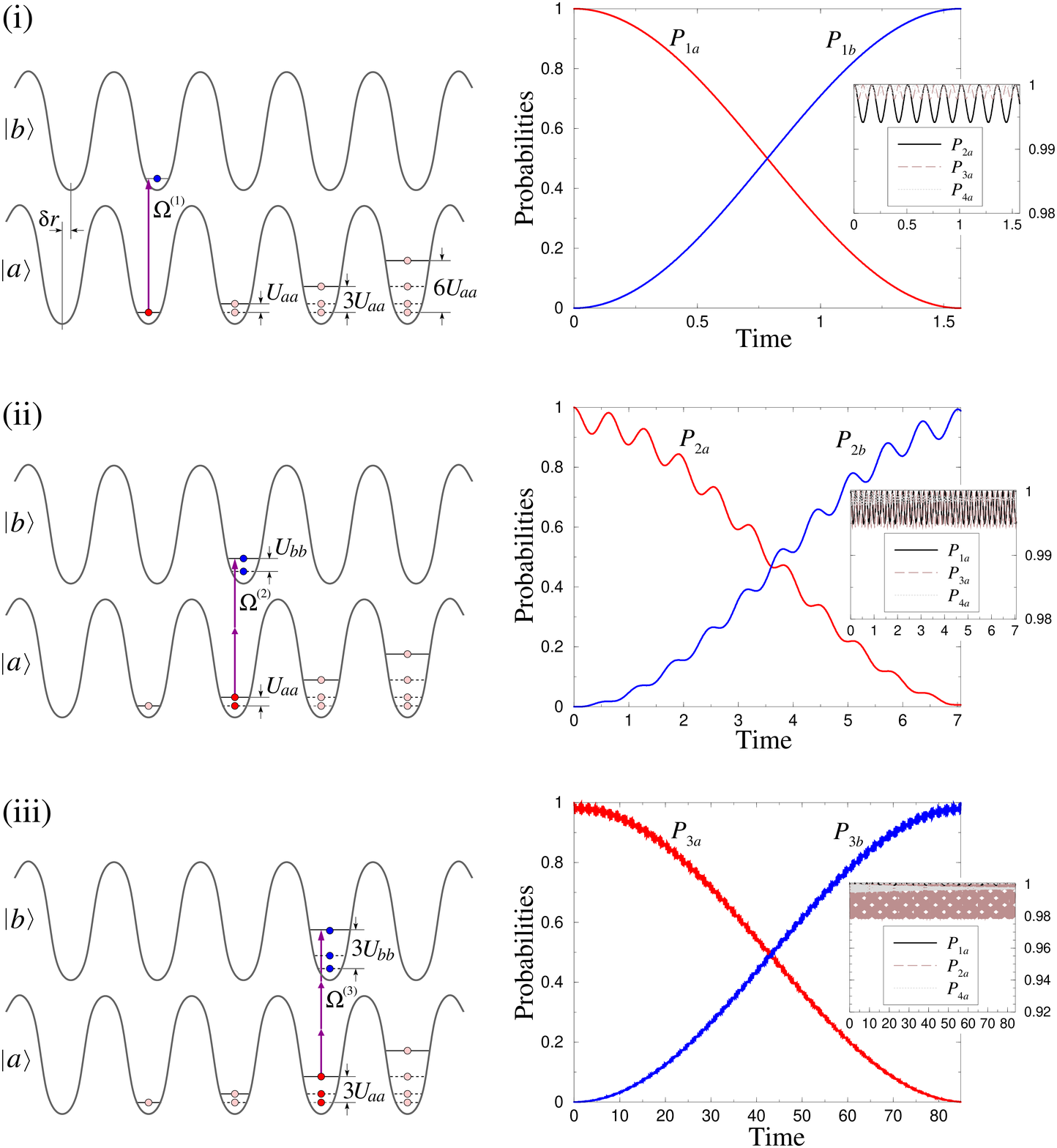}}
\caption{Schematics (left column) and dynamics (right column) 
of atom transfer between two optical lattices: 
selective transfer of 
(i) single atoms, $N=1$; (ii) pairs of atoms, $N=2$; and 
(iii) triples of atoms, $N=3$, via the corresponding $N$-photon 
resonant transitions. The main graphs display the probabilities 
$P_{N_{\alpha}}$ for $N$ atoms in the corresponding states 
$\ket{\alpha}$ ($\alpha = a,b$), while the insets show the
probabilities $P_{n_a}$ of initial states $\ket{n_a}$ with $n \neq N$.
(For $n>4$, the probabilities $P_{n_a}$ oscillate 
with even smaller amplitudes and therefore not shown.)
The numerical simulations employ the parameters $U_{\alpha \alpha'}$ 
and $\Om$ listed in the text, the time is in units of $\Om^{-1}$
and the evolution terminates at the corresponding $\tau^{(N)}$.} 
\label{fig:poptr}
\end{figure*}

We now discuss the transfer of selected number of atoms $N$ between 
the two lattices, as illustrated in Fig.~\ref{fig:poptr} (left panel). 
This atom-number filtering procedure is very simple yet remarkably 
efficient and robust, provided 
\besa 
|U_{ab} - \frac{1}{2}(U_{aa} + U_{bb})| \gg \hbar \Omega , \\
|U_{aa} - U_{bb}|, |U_{ab} - U_{aa,bb}| \gg \hbar \Omega .
\eesa 
The first of these conditions ensures that within the selected 
$N$-atom subspace all the intermediate states are nonresonant,
while the remaining conditions are needed to suppress all the 
transitions out of the other initial states $\ket{n_a}$ with
$n \neq N$, as clarified below. For convenience we denote 
$\de U \equiv [U_{ab} - \frac{1}{2}(U_{aa} + U_{bb})]/\hbar$. 

%%%%%%%%%%%%%%%%%%%%%%%
%%%%%%%%%%%%%%%%%%%%%%%

\paragraph*{(i) Single atom transfer, $N =1$.}

To filter out only the single atoms per site, we tune the frequency 
of the coupling field to be resonant with the atomic 
transition $\ket{a} \to \ket{b}$, i.e., we set $\om = \om_{ba}$. 
The field will then induce resonant Rabi oscillations between the 
states $\ket{1_a,0_b}$ and $\ket{0_a,1_b}$ with frequency $\Om^{(1)} = \Om$. 
If we apply the field for time $\tau^{(1)} = \pi / 2 \Om^{(1)}$, 
resulting in a $\pi$-pulse, all the single atoms $\ket{1_a}$ will 
be transferred to $\ket{1_b}$.

\paragraph*{(ii) Two atom transfer, $N=2$.}

To filter out only the pairs of atoms per site, we choose 
the frequency of the coupling field according to the condition 
$2 \om = 2\om_{ba} + (U_{bb} - U_{aa})/\hbar$, which implies a 
two-atom (and two-photon) transition 
$\ket{2_a,0_b} \to \ket{1_a, 1_b} \to \ket{0_a,2_b}$ via 
nonresonant intermediate state $\ket{1_a, 1_b}$ detuned by $\de U$. 
The corresponding two-atom (-photon) Rabi frequency is then 
$\Om^{(2)} = 2 \Om^2 /\de U$ (the factor of $2 =2!$ originates 
from double application of bosonic operators $\hbd \ha$ to the initial 
state $\ket{2_a,0_b}$), and at time $\tau^{(2)} = \pi / 2 \Om^{(2)}$,
corresponding to an effective $\pi$-pulse, all the pairs of atoms 
$\ket{2_a}$ will be transferred to $\ket{2_b}$.

\paragraph*{(iii) Three atom transfer, $N=3$.}

To filter out only the triples of atoms per site, 
we choose the frequency of the coupling field according to
the condition $3 \om = 3 \om_{ba} + 3 (U_{bb} - U_{aa})/\hbar$,
which implies a three-atom (-photon) transition 
$\ket{3_a,0_b} \to \ket{2_a, 1_b} \to \ket{1_a, 2_b}  \to \ket{0_a,3_b}$ 
via nonresonant intermediate states $\ket{2_a, 1_b}$ and $\ket{1_a, 2_b}$ 
both detuned by the equal amount of $2 \de U$. The corresponding 
three-atom (-photon) Rabi frequency is then 
$\Om^{(3)} = 6 \Om^3 /(2 \de U)^2$ (the factor of $6 = 3!$ originates 
from triple application of $\hbd \ha$ to state $\ket{3_a,0_b}$). Note 
that since both intermediate states $\ket{2_a, 1_b}$ and $\ket{1_a, 2_b}$
have the same detuning $2 \de U$, the second-order ac Stark shifts of 
states $\ket{3_a,0_b}$ and $\ket{0_a,3_b}$ are the same, given by 
$3 \Om^2/(2 \de U)$, and the differential shift on the three-photon 
transition $\ket{3_a,0_b} \to \ket{0_a,3_b}$ vanishes. 
Hence, applying the field for time $\tau^{(3)} = \pi / 2 \Om^{(3)}$, 
corresponding to an effective $\pi$-pulse, all the triples of atoms 
$\ket{3_a}$ will be transferred to $\ket{3_b}$.

The above procedure can be generalized to multiphoton transfer 
of any number of atoms $N$ between the two lattices.
Under the $N$-photon resonance condition 
$N \om = N \om_{ba} + \frac{1}{2}N(N-1) (U_{bb} - U_{aa})/\hbar$, 
the effective $N$-atom (-photon) Rabi frequency is then given by
\[
\Om^{(N)} = \frac{ N! \, \Om^N}{ [(N-1)!]^2 \de U^{N-1}} 
= \frac{ N \Om}{ (N-1)! } \left( \frac{\Om}{\de U} \right)^{N-1} .
\]
However, due to the above scaling of $\Om^{(N)}$ and condition 
$\Om \ll |\de U|$, the corresponding transfer time 
$\tau^{(N)} = \pi/2 \Om^{(N)}$ will become prohibitively long for 
$N \geq 4$ in a realistic optical lattice experiment, as discussed below.  

In Fig. \ref{fig:poptr} (right panel) we demonstrate, via numerical 
solution of the corresponding Schr\"odinger equations, that the 
transfer  of the selected number of atoms $N=1$, $2$ and $3$ between 
the two lattices is indeed very efficient, with the probabilities 
$P_{N_b} (\tau^{(N)})$ of the final states $\ket{N_b}$ at the corresponding 
times $\tau^{(N)}$ being close to unity, while the probabilities
$P_{n_a}$ of initial states $\ket{n_a}$ with $n \neq N$ changing 
very little during the transfer. For these simulations, we choose 
$U_{aa}/\hbar \simeq 2 \pi \times 10^4\:$s$^{-1}$, and, 
upon assuming $a_{aa} \simeq a_{bb} \simeq a_{ab}$, $V_a/V_b =3$ and 
$\de r =0$, obtain from Eqs.~(\ref{Uaap}) $U_{bb} \simeq 0.44 U_{aa}$
and $U_{ab} \simeq 0.63 U_{aa}$. We then have 
$\de U \simeq 2 \pi \times 930\:$s$^{-1}$
and set $\Om = 2 \pi \times 100\:$s$^{-1}$. The corresponding 
one-, two- and three-atom transfer times are given, respectively, by 
$\tau^{(1)} = 2.38 \times 10^{-3}\:$s, $\tau^{(2)} = 1.16 \times 10^{-2}\:$s
and $\tau^{(3)} = 0.144\:$s, which are shorter than the typical lifetimes
($0.5\:$s) of cold atoms in optical lattice MI shells with $n \leq 3$  
\cite{n12345}.

%\section*{Applications and conclusions}

Hence, using our procedure one can separate the spacial MI shells 
of the optical lattice with different atom numbers $n$ \cite{n12345,n12}, 
placing in another lattice only the desired $N$th shell, which can 
be a filled sphere (or circle in 2D), or a hollow sphere (ring in 2D), 
depending on whether it is extracted from the central part of the 
trap or not. This is then followed by discarding (releasing) 
the atoms of the first lattice. 
Another useful application of our atom-number filtering technique 
is a preparation of pure samples of the interaction-bound lattice dimers 
\cite{KWEtALPZ,molmer,MVDPdmrs,DPKLT}, or trimers \cite{MVDPAZ},
without resorting to more complicated procedures involving Feshbach 
association, purification and dissociation of atom pairs 
\cite{KWEtALPZ,feshmols}. 

%%%%%%%%%%%%%%%%%%%%%%%
\begin{figure}[t]
\centerline{\includegraphics[width=0.48\textwidth]{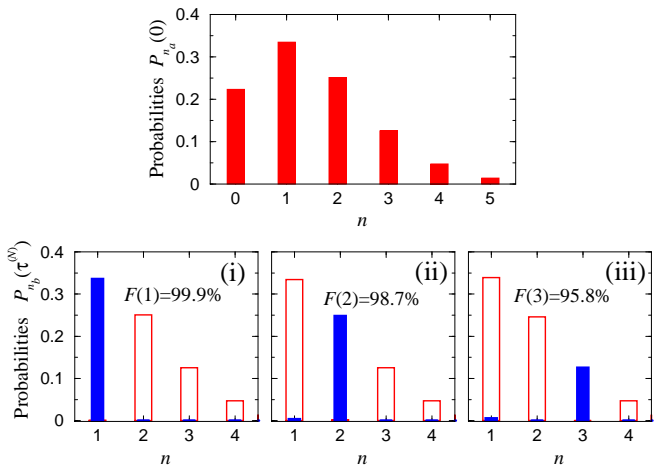}}
\caption{Filtering out of the initial Poisson distribution with mean
atom number $\expv{n}_a =1.5$ (top panel graph) single atoms (i), 
pairs of atoms (ii), and triples of atoms (iii). In the lower 
panel graphs, the empty (red) bars show the atom-number probability 
distributions $P_{n_a}(\tau^{(N)})$ remaining in the first lattice 
after the transfer. The transfer fidelities $F(N)$ are defined in
Eq. (\ref{FidN}). All parameters are as in Fig.~\ref{fig:poptr}.}
\label{fig:chrflt}
\end{figure}
%%%%%%%%%%%%%%%%%%%%%%%

As an example, in Fig.~\ref{fig:chrflt} we illustrate the 
filtering out of the initial Poisson atom-number distribution 
$P_{n_a}(0) = \expv{n}^n e^{-\expv{n}}/n!$, corresponding to a frozen 
SF phase with mean occupation number $\expv{n}_a =1.5$, the desired 
number of atoms $N=1$, $2$ or $3$. The only variables adjusted to 
each $N$ case are the coupling field frequency $\om$ and transfer 
time $\tau^{(N)}$, with all the other parameters the same, as described 
above in connection with Fig. \ref{fig:poptr}. We quantify the transfer
using the fidelity
\be
F(N) = \frac{P_{N_b}(\tau^{(N)})}{\sum_{n>0} P_{n_b}(\tau^{(N)})} , \label{FidN}
\ee
for which we obtain very high values $F(1) = 0.999$, $F(2) = 0.987$ and
$F(3) = 0.958$. 

To conclude, we have proposed and analysed a very efficient and robust 
procedure to filter out from an optical lattice with an arbitrary 
inhomogeneous site occupation number only preselected number of bosonic 
atoms per site and place them into another internal atomic state, 
creating thereby a lattice with desired site occupation number, 
which we envision to have a number of interesting applications.

%\begin{acknowledgments}
%This work was supported by the EC Marie Curie Research-Training 
%Network EMALI (MRTN-CT-2006-035369). 
%\end{acknowledgments}

\end{document}